\documentclass[10pt,twocolumn,showpacs,preprintnumbers,amsmath, amssymb, amsfonts,prb,aps]{revtex4-1}
\usepackage{articlestyle}
\newcommand\ie {{\it{i.e.}}}
\newcommand\eg {{\it{e.g.}}}

\begin{document}
\title{The quantum Hall hierarchy in spherical geometry}
\author{Thomas Kvorning}
\email[E-mail: ]{thomas.kvorning@fysik.su.se}
\affiliation{%
Department of Physics, Stockholm University, AlbaNova University Center, SE-106 91 Stockholm, Sweden  \\
}%
\begin{abstract}
Representative wave functions, which encode the topological properties of the spin polarized fractional
quantum Hall states in the lowest Landau level, can be expressed in terms of  correlation functions in conformal 
field theories. Until now, the constructions have  been restricted to flat geometries, but in
 this paper we generalize to the simplest curved geometry, namely that of a sphere.
  Except for being of interest for numerical studies, that usually are performed on a sphere, the response 
  of the FQH liquids  to curvature can be used to detect a topological quantity, the shift, $\mathcal S$, which is
  the average orbital spin of the constituent electrons. 
 We give explicit expressions for representative  wave functions on the sphere, for the full Abelian FQH hierarchy, and
 calculate the corresponding shifts. These  microscopic  results,  based on wave functions,  agree with the predictions from the effective Chern-Simons field theory. 
 The methods we develop can also be applied to the planar case. It  gives simpler expressions for  states 
with both quasiparticle and quasihole condensates, and  allows us to give closed form expressions for a general 
state in the hierarchy, rather than finding the wave function on a case by case basis.

\end{abstract}

\maketitle

\section{Introduction}
Most of the observed quantum Hall states in the lowest Landau level (LLL) can be understood as part of the Haldane-Halpering hierarchy\cite{haldane83, halperin84}, where the daughter of a parent state is obtained by condensation of quasiparticles or quasiholes into a correlated state of the Laughlin type\cite{laughlin83}. 
In its original formulation, the resulting wave functions were complicated since they involved multi-dimensional integrals at each level of the hierarchy, but in recent work, based on the close connection between QH wave functions and correlators in certain conformal field theories (CFTs) \cite{moore91},  a simpler picture  has emerged.  
In Refs. \onlinecite{hansson07, hermanns08, bergholtz07, *bergholtz08, hansson09a, *hansson09b},  representative wave functions for all quasiparticle condensates in the hierarchy were constructed, and in two more recent papers \cite{suorsa11a, suorsa11b} this work was extended to include the full spin polarized hierarchy.  

The main reason for expressing QH wave functions in terms of conformal blocks, is that the topological properties of electrons and quasiparticles are conjectured to be related to charges and conformal spins of the corresponding operators.\footnote{
 To prove this conjecture, one must show that the phase factors that determine charge and statistics of the quasiparticles, can be read directly form the monodromies of the pertinent conformal blocks, without any additional Berry phases. This has been demonstrated in several cases\cite{read96, bonderson11}, and we shall assume it to be true also for the hierarchy states although here a  proof is so far lacking.
 }
Using this conjecture as a working hypothesis gives two motives to why the conformal blocks are used as wave functions: Firstly, the wave functions can be used to see how topological properties are encoded in electron states, and how microscopic state supporting anyons manifest themselves. Secondly, by comparing numerical studies with these wave functions more information about which topological states that are realized by some realistic interaction can be obtained.
 
In this paper, we extend the techniques in Refs. \onlinecite{hansson07}-\onlinecite{suorsa11b}, to spherical geometry. There are many reasons for doing this. One is an important practical reason: most numerical tests of QH wave functions are done on the sphere, and it is thus important to have concrete expressions for the corresponding wave functions. So far, most numerical work on hierarchy wave functions has been restricted to  composite fermion states\cite{jain07}, while we here will provide techniques to construct explicit wave functions for an arbitrary state in the hierarchy. To connect to previous work,  we also show that at the Jain filling fractions $\nu = n/(2pq\pm 1)$, our  wave functions are identical to those obtained using composite fermions. 

There are also theoretical reasons for generalizing to the spherical geometry. One is that the sphere provides the simplest setting for studying how the hierarchical QH liquids respond to curvature. This was investigated early on by Wen and Zee, in the context of an effective Chern-Simons description of the Abelian QH states\cite{wen92b}. The shift, $\mathcal S$, is the offset between the number of flux quanta, $N_\phi$ penetrating the sphere and the number, $N/\nu$, expected from analogy with the plane. Wen showed that it is a topological rather than geometrical quantity, and  thus a proper part of the topological characteristic of the state. Read\cite{read09} has given general arguments for  the shift to equal half the average conformal (or orbital) spin, and our explicit calculations verify this for all states in the hierarchy. 

The paper is organized as follows. In section II and III we introduce the conformal block wave functions on the plane to give some technical background and to introduce  notation. 
We also give a systematic discussion of the various constraints, due to symmetry and regularity, that must be imposed on the CFT operators describing the electrons. It turns out that some constraints that were implicit in the earlier treatments are not necessary, and are in fact difficult to impose on the sphere. 
We give alternative explicit expressions that treat quasiparticles and quasiholes in a symmetric way, and can conveniently be used in the spherical geometry. This choice  is, however, not unique, but leaves room for short distance modifications, which are  necessary in order to find the actual ground-state for  a realistic Hamiltonian.   
 The later sections, section IV - VII, contain our main new results on hierarchy wave functions on the sphere. Section VIII offers a summary of our results and some ideas for future work.  
\section{Preliminaries}
In this and the next section we will review the construction of representative electronic wave functions for arbitrary hierarchical states, with a topological long range behavior described by a Chern-Simons (CS)  Lagrangian of the type described in  Refs. \onlinecite{wen92b} and \onlinecite{wen95,*wen92a}  (rational quantum Hall (RQH) states).
These wave functions are expressed as linear combinations of  correlators of operators in certain CFT’s, so our first task is to identify the proper CFT and the pertinent operators that correspond to the CS theory in question. 

The techniques in this paper can be used for any RQH state, but our main focus is the spin-polarized, single layer, Haldane Halperin hierarchy. The idea of the Haldane-Halperin hierarchy is that just as electrons can “condense” into a Laughlin state, the quasiparticles or quasiholes of an arbitrary QH “parent state” can condense to form a “daughter state”. The Laughlin states together with all the daughter states will form the full QH hierarchy of odd denominator states. 

The effective CS theory determines all topological properties of the RQH liquids: the Hall conductance and viscosity, the degeneracy on higher genus surfaces, and the topological quantum numbers of the elementary excitations, \ie, their charge, spin,  and  braiding statistics. More precisely, the CS action is expressed in terms of an $n\times n$ integer K-matrix $\mathbf K$, a  charge vector, $\mathbf t$, and a  spin vector $\mathbf s$. In a general hierarchy state, there are many distinct elementary excitations, each characterized by an integer vector $\mathbf l$. The  charge $q_{\mathbf l}$, orbital spin $s_{\mathbf l}$, and phase $\theta_{\mathbf l,\mathbf l'}$ obtained by braiding it around some other excitation $\mathbf l'$ is given by,
\alr{
	q_{\mathbf l}&=\mathbf l^\tr \mathbf K^{-1} \mathbf t &
	s_{\mathbf l}&=\mathbf l^\tr \mathbf K^{-1} \mathbf s &
	\theta_{\mathbf l,\mathbf l'}&=2 \pi \mathbf l^\tr \mathbf K^{-1}\mathbf l' \ .  \label{qqnumbers}
} 
The topological properties of the liquid  does not uniquely determine the CS action since two triplets $(\mathbf K^\prime, \mathbf t^\prime, \mathbf s^\prime)$ and $(\mathbf K, \mathbf t, \mathbf s)$ related by $(\mathbf K^\prime, \mathbf t^\prime, \mathbf s^\prime)= (\mathbf {WKW}^T, \mathbf W \mathbf t, \mathbf W \mathbf s)$, where $\mathbf W\in GL(n,\mathbb Z)$, are equivalent. It is, for instance, easy to see that the quantum numbers in (\ref{qqnumbers}) stay the same if we simultaneously change $\mathbf l$ to $\mathbf W^{-1} \mathbf l$.

For the Haldane-Halperin hierarchy there is a natural representation for the K-t-s triplet: If the condensation occurs $n$ times we can think of the state as composed by $n$ fictitious layers, and if we let $2 \pi K_{IJ}$ denote the braiding phase gotten when one electron in layer $I$ braid around one in layer $J$, we get the representation
  \al{
		K_{IJ}&=\gamma_{I}(\delta_{IJ}+1)+  \mspace{-20mu}\sum_{k=0}^{\min(I,J)-1}\mspace{-20mu} \gamma_k p_k-2 \gamma_{k+1}\\
		t_I&=1 \mspace{67mu} s_I=\frac12 K_{II}+\sum_{k=1}^{I}\gamma_k \ .
	}
We have here used Wen's notation\cite{wen95}, \ie, $p_k$ denotes the filling of the $k$'th condensate%
 \footnote{$p_0=m$ in Haldane’s  original paper\cite{haldane83}}
 and $\gamma_k = \pm 1$ depending on whether the condensed quasiparticles in the $k$'th condensate have  the same ($+$) or opposite ($-$) charge as the electron.

The CFT we will use, can be formulated in terms of massless bosons denoted by  $\varphi,\bar\varphi,\phi$, and an action where each boson has a term normalized as 
	\alr{
		S[\varphi]=-\frac1{8\pi}\int d^2x\sqrt g \varphi \Delta \varphi  \ .
		\label{action}
	}
	Here $\Delta$ is the Laplace operator, and $d^2x \sqrt g$ is the surface element. In cartesian coordinates we have 
$\Delta = \pd_\mu\pd^\mu$, and $\sqrt g\equiv 1$. We use the convention that $\mathbf x=(x,y)$ are the cartesian coordinates on the plane, or isothermal coordinates on the sphere. Cartesian coordinates are defined by, $g_{\mu\nu}(\mathbf x)=\delta_{\mu\nu}$, and isothermal coordinates are a generalization where $g_{\mu\nu}(\mathbf x) = f(\mathbf x)\delta_{\mu\nu}$ for some function $f$. 

To properly define the CFT  it is not enough to give the action, but we must also specify the operator content. The theories relevant for the RQH are 
rational CFTs, and for bosonic theories this means that  the fields must be compactified on a lattice. As a consequence, all operators carry integer charges with respect to the conserved currents $J^i_\mu = \frac 1{i\pi}\pd_\mu \varphi_i$, related to the  the  $U(1)$ symmetries, $\varphi_i\rightarrow\varphi_i+a_i$. In the next section we will show how the charge lattice is determined from the K-matrix. 

 According to the Moore-Read conjecture the QH wave functions are to be identified with the conformal blocks of the relevant CFT. These can be extracted either by factorizing the full correlation functions in chiral parts, which is always possible in a rational CFT, or by directly calculating correlation functions of chiral operators. In either case, there is a freedom that amounts to a re-phasing of the wave function (\ie a gauge transformation). %
This choice has no physical significance. A technically simple choice, which also is used in the earlier papers, is to define the chiral parts of the bosons by removing all positive or negative eigenstates of the operator $L_3=z\pd-\bar z\bar\pd$; this will result in wave functions in the symmetric gauge. Here $\pd\equiv \frac{\pd}{\pd z}$, $\bar z=x-i y$ and $\bar\pd\equiv \frac{\pd}{\pd \bar z}$. We will from now on use this notation so that \eg $\bar\partial_i$ mean $\partial/\partial\bar z_i$. Thus, a chiral boson is defined as  $\varphi_{R/L}(\mathbf x)=P_{R/L}\varphi(\mathbf x)$, where $P_{R/L}$ is the projection operator onto the space of non-negative or non-positive eigenstates of $L_3$, respectively.

 The action (\ref{action}) is normalized to give the two-point-function
	\al{
		\braket{\varphi(\mathbf x_1)\varphi(\mathbf x_2)}=\ln|\mathbf x_1-\mathbf x_2|^2=\\
		=\ln(z_1-z_2)+\ln(\bar z_1-\bar z_2) \ ,
	}
	 and the chiral and anti-chiral two-point-functions are, up to an additive constant, 
	 \al{
	 	\braket{\varphi_L(\mathbf x_1)\varphi_L(\mathbf x_2)}&=\ln(z_1-z_2) \\
		\braket{\varphi_R(\mathbf x_1)\varphi_R(\mathbf x_2)}&=\ln(\bar z_1-\bar z_2) \ .
	 }
	Notice that we here, and in the rest of the article, will use $\mathbf x$ as the argument for operators altough they are ``almost analytic'' in $z$. 	
\section{The FQH hierarchy on the plane}
\subsection{The Laughlin state}	 
The K-t-s triplet for the $\nu = 1/m$ Laughlin  state are numbers $(\mathbf K,\mathbf t,\mathbf s)=(m, 1,\frac m2)$, and  the topologically distinct excitations are labeled by an integer $l$.   
The chiral vertex operators
	\al{
		\mathscr O_l(\mathbf x)  =\noeq e^{-i l/\sqrt m \varphi_R}\no (\mathbf x)\ ,
	}
have the property
	\alr{
		\mathscr O_l(\mathbf x)\mathscr O_{l'}(\mathbf x')=(z-z')^{l l'/m}\no\mathscr O_l(\mathbf x)\mathscr O_{l'}(\mathbf x')\noeqr \ . \label{opel}
	}
	Here $:\cdots:$ denotes normal ordering of the bosons, by point-splitting all pairs and subtracting the correlator of the pair.
	
	To use these operators to obtain an electronic wave function we must identify the proper electron operator  by demanding that it has  unit charge and also  trivial braiding  relative to all other particles. This is achieved by taking\footnote{The electron operators which make up the background will have conjugated $U(1)$ charge with respect to electron excitations, see Ref. \onlinecite{hansson07}.}
	 $l_{el}=-m$, and the corresponding chiral vertex operator is 
	\al{
		V(\mathbf x)=\noeq e^{i\sqrt m \varphi_R}\no (\mathbf x)\ .
	}
The correlator of $N$ of these operators will give the $N$-particle Laughlin wave function. This follows since \eqref{opel}  gives the holomorphic part of the Laughlin wave function. We can also insert fundamental quasiholes with the operator $H(\mathbf x)=\noeq e^{i/\sqrt m \varphi_R}\no (\mathbf x)$, and then we get the following expression for the $N$ electron system with $N_h$ quasiholes, 
	\mlr{
		\braket{\prod_{i=1}^N V(\mathbf x_i) \prod_{i=1}^{N_h} H(\mathbf x^h_i)}=\\
		=e^{-\frac1{4l^2}\sum_i |\mathbf x_i|^2}e^{-\frac{|q_h|}{4l^2}\sum_i |\mathbf x_h|^2}(1-h)(1-1)^3(h-h)^{1/3} \ .
		\label{Laughlinwf}
	}
	In the equation above $q_h$ denotes the quasihole charge in units of the electron charge, in this case $q_h=-\frac13$. We also use the short hand notation $(1-1)\equiv\prod_{i<j}(z_i-z_j)$, $(h-h)\equiv\prod_{i<j}(z^h_i-z^h_j)$ and $(1-h)\equiv\prod_{i,j}(z_i-z^h_j)$. 

The Gaussian factors in \eqref{Laughlinwf} require an explanation. Correlators of this type, where the $U(1)$ charges do not sum to zero, vanish identically,
which  is seen most easily  using path integrals. Following Ref. \onlinecite{hermanns08}  we use the idea from Ref. \onlinecite{moore91} and remedy this by supplementing the action with a neutralizing background field,  
\al{
		S\rightarrow S -\frac i{2\pi \sqrt m }\int  \mathcal B\p{ \varphi^\alpha} \ ,
	}
where $\mathcal B$ is  the magnetic field multiplied by $2\pi$ and divided by the magnetic flux quantum. With these conventions the integral of $\mathcal B$ is the number of flux quanta ($N_\phi$) multiplied by $2\pi$, \ie, $N_\phi=\int\mathcal B/2\pi$. The normalization is chosen so as to cancel the total $U(1)$ charge of the operators. For the ground state this implies  the relation $N=\mathrm K^{-1} N_{\phi} = N_\phi /m$,  between the number of electrons and flux quanta,  which follows from the effective CS theory.  With $N_h$ quasiholes present this condition is modified to the relation $N=\frac1m N_\phi- \frac1m N_h=\frac1m N_\phi+q_h N_h$. Also, a careful evaluation of the correlator using a proper regularization will  also produce the Gaussian factors in \eqref{Laughlinwf}. For details, we refer to Ref. \onlinecite{hermanns08}.
\subsection{The chiral sector}
	Before turning to a completely general state in the hierarchy, we generalize the above discussion to the fully chiral part of the hierarchy. This is obtained by consecutive condensations of quasiparticles only, \ie  by excitations with the same charge as the electron. At  level $n$, there are sets $S_{el}=\{l_e^i\}_{i=1,\dotsc n}$ of $n$ linearly independent excitations, all  with the quantum number of electrons, and the ground state wave function can be constructed from correlators of these operators. 
In the multilayer representation one such set can be formed by $S_{el}=\{-\mathbf K \mathbf e_i\}_{i=1,\dotsc n}$ where $\mathbf e_i$ is the unit vector. Just as with the Laughlin case the relevant primary chiral operators are formed from the square root of the K-matrix, \ie as  
		\al{
			V_I&=\pd^{I-1}\no e^{i Q_{IJ}\varphi^J_R}\no & I=1,\dotsc n \ ,
		}
		with
		\al{
			\mathbf K&=\mathbf Q \mathbf Q^T \ .
		}
		Here $\varphi^1,\varphi^2,\dotsc$ are  bosons with the action (\ref{action}), and summation over an upper and a lower index should be understood. From now on, $\mathbf K$ always denotes the $K$-matrix in multilayer representation. The derivatives ensure that the operators carry the correct orbital spin, which \eg encode the Hall viscosity (see Ref. \onlinecite{read11}), as discussed in  in Ref. \onlinecite{suorsa11a}. 
		As in the Laughlin case we have to add a background term to the action in order not to get vanishing correlators,
		\al{
			S\rightarrow S -\frac i{2\pi}\int  \mathcal B c_I \varphi^I \ .
		}
		The neutrality condition now becomes 
		\alr{
			N_\phi t_I=\sum_J K_{IJ} N_J\ ,
			\label{neutralitywen}
		}
		 and to satisfy it, we  take $\mathbf c =\mathbf Q^T \mathbf K^{-1}\mathbf t $.
		The ground state  wave function for $N$ particles is now obtained from 
		\al{
			\Braket{V_1(\mathbf x_1)\dotsb V_1(\mathbf x_{N_1})V_2(\mathbf x_{N_1+1})\dotsb V_n(\mathbf x_N))} \ .
		}
		The correlation function vanishes unless it is neutral with respect to the $U(1)$ charges. This implies the relations
		\alr{
			\sum_I N_I Q_{IJ}&=N_\phi c_J & J=1,2,\dotsc \ ,
		}
		which is equivalent to \eqref{neutralitywen}. When the relation is fulfilled, the correlation function is proportional to
		\alr{
			e^{-\frac1{4l^2}\sum_i |\mathbf x|^2}(1-1)^{K_{11}}\pd_2(1-2)^{K_{12}}(2-2)^{K_{22}}\pd^2_3\times\dotsb
			\label{coherentstatewavefunction}
		}
		where we used a notation equivalent to that in \eqref{Laughlinwf}. Here the numbers $1,2,\dotsc$ denote which electron operator the coordinate correspond to, and $\pd_I$ denote a derivative with respect to all coordinates corresponding to the $I$'th electron operator.
		
		The above expression is, however, not an acceptable wave function, since it is not anti-symmetric in the electron coordinates. Rather than directly anti-symmetrizing this expression we use a more general formalism, valid for the full hierarchy, which we also will use later in the case with spherical geometry. 
		The basic idea, explained in detail in  Ref. \onlinecite{suorsa11a},  is to regard \eqref{coherentstatewavefunction} as a coherent state wave function. To get the wave function in the position basis, we convolute it with a coherent state kernel as
		\mlr{    \label{conv}
		\Psi(\{\xi_i\})=\int \prod_i d^2 x_i \braket{\xi_1,\dotsc, \xi_N|z_1,\dotsc, z_N}\times\\\times \Braket{V_1(\mathbf x_1)\dotsb V_1(\mathbf x_{N_1}) V_2(\mathbf x_{N_1+1})\dotsb V_2(\mathbf x_{N})} .
		}
The kernel is given by
		\alr{
			\braket{\xi_1,\dotsc, \xi_N|z_1,\dotsc, z_N}
			\propto\mathcal A\prod_i e^{-1/{4l^2}(|z_i|^2-2\xi_i \bar z_i+|\xi_i|^2)} \ .	
		}
		As earlier $z_i$ denote $z_i\equiv x_i+iy_i$ where $(x_i,y_i)\equiv\mathbf x_i$ and $\mathcal A$ denotes anti-symmetrization of the electron coordinates. In this case the convolution amounts to nothing but an anti-symmetrization, but in general this is not true. Using the coherent state kernel we will, for all wave functions considered in this paper, always be able to perform the integrals exactly, to get  explicit closed form expression for the position basis wave functions.
		
		\subsection{The full hierarchy}
	For a general hierarchy state, that also involves condensation of quasiholes, the K-matrix  is no longer positive definite, and cannot be written as $\mathbf K=\mathbf Q \mathbf Q^T$.
	In Refs. \onlinecite{suorsa11a, suorsa11b} this problem is resolved by introducing a  two-component picture, obtained by splitting the K-matrix into a chiral and anti-chiral part $\mathbf K={\boldsymbol\kappa}- \bar{\boldsymbol\kappa}  =    \mathbf Q \mathbf Q^T -  \bar{\mathbf Q} \bar{\mathbf Q}^T  $,
and defining the electron operators,
\al{
		V_I=\pd^{\sigma_I}\bar\pd^{\bar\sigma_I}\no e^{iQ_{IJ}\varphi_L^J+\bar Q_{IJ}\bar \varphi_R^J}\noeqr \ .
	}
Since we have both chiralities, we can have both holomorphic and anti-holomorphic derivatives. To get correct
orbital spins, we must have
\alr{
		\sigma_I-\bar\sigma_I+\frac12 K_{II}=s_I \ .   \label{spinvalue}
	}
	Note that the operators are in no way unique. The CS theory only imposes constraints on the differences $\sigma_I-\bar\sigma_I$ and ${\boldsymbol\kappa}-\bar{\boldsymbol\kappa} $. The different choices is reflected in the short-distance behavior of the wave function. 

Finally, to get the correct number of particles we must introduce a background term for both chiralities,
	\al{
		S\rightarrow S -\frac i{2\pi}\int  \mathcal B\p{ c_I \varphi^I+\bar c_I \bar\varphi^I} \ ,
	}
and set $\mathbf c=\mathbf Q^T \mathbf K^{-1}\mathbf t $ and $ \mathbf {\bar c}=\mathbf {\bar Q}^T \mathbf K^{-1}\mathbf t$. 
\subsection{The charge lattice}
Just as in the Laughlin case, the fundamental excitations carry fractional charge with respect to the electron. In the CS theory the allowed excitations are described by  integer $\mathbf l$ vectors, and the  corresponding CFT operators are given by 
	\alr{
		\mathscr O_{\mathbf l}(\mathbf x)   = \noeq  e^{-i(r_{\mathbf l})_I\varphi^I_R-i(\bar r_{\mathbf l})_I\bar \varphi^I_L}\noeqr \ ,
		\label{excitations}
	}
	with
	\al{
		\mathbf r_{\mathbf l}&=\mathbf Q^T \mathbf K^{-1} \mathbf l & \bar{\mathbf r}_{\mathbf l}&=\bar{\mathbf Q}^T \mathbf K^{-1} \mathbf l \ .
	}
These  primary operators determines the charge lattice up to a normalization of the currents which we take as
	\al{
		J^I_\mu&=\frac{1}{i \pi}\pd_\mu\varphi^I& \bar J^I_\mu&=\frac{1}{i \pi}\pd_\mu\bar\varphi^I \ .
	}
	 With this normalization  the charge lattice becomes
	\al{
		\bigl\{(\mathbf r,\bar{\mathbf r})\bigr|(\mathbf r,\bar{\mathbf r})=\mathbf Q^T \mathbf K^{-1} \mathbf l\oplus\bar{\mathbf Q}^T \mathbf K^{-1} \mathbf l \text{ with } l_I\in\mathbb Z\bigr\} \ .
	}
	\section{Explicit choice of vertex operators}
	\label{convention}
	We already stressed that, given a CS theory, there are many choices of CFT operators representing electrons and quasiparticle excitations that will give wave functions with identical topological characteristics but different short-distance behavior. This freedom is necessary, since small changes in the Hamiltonian will change the wave functions, but not the topological properties. Within the class of wave functions obtained in the CFT scheme described above, the short-distance freedom is related to exactly where the derivative act, and to the possibilities of adding extra factors of the form $\prod_{ij} |z_i - z_j|^{2p}$, that increase the repulsion between particles without changing the angular momentum. It turns out that not all the possibilities on the plane can be realized on the sphere. In  this section we first discuss the most general form of the hierarchical vertex operators, and then introduce a  particular choice that  does carry over to the sphere, a choice which also is advantageous to use on the plane for reasons to be explained below. 
	
		In the hierarchal construction of the electron operators (see Refs. \onlinecite{hermanns08}-\onlinecite{hansson09a, *hansson09b}) the $\kappa$-matrices take a special form since the daughter states inherit certain structures from the parent state. Concretely, this is manifested as
\begin{align}
	\begin{pmatrix}
		\kappa_{11}& \cdots&\kappa_{1n}\\
		\vdots &\ddots&\vdots \\
		\kappa_{n1}&\cdots&\kappa_{nn}
	\end{pmatrix}\rightarrow
		\begin{pmatrix}
		\kappa_{11}&\!\!\!\cdots\cdots&\!\!\!\!\kappa_{1n}&\!\!\kappa_{1n}\\
		\text{\rotatebox{90}{$\, .\, .\, .\, .\, .\, .\, .$ }} &\text{\rotatebox{135}{$\, .\, .\, .\, .\, .\, .\, .\, .\, .$ }}&\!\!\!\!\!\!\text{\rotatebox{90}{$\, .\, .\, .\, .\, .\, .\, .$ }} &\!\!\text{\rotatebox{90}{$\, .\, .\, .\, .\, .\, .\, .$ }}  \\
		\kappa_{n1}&\!\!\!\cdots\cdots&\!\!\!\!\!?&\!\!?\\
		\kappa_{n1}&\!\!\!\cdots\cdots&\!\!\!\!\!?&\!\!?
	\end{pmatrix}
	\label{matrix}
\end{align}
which shows the transition from a parent state at level $n$ to a daughter at level $n+1$. Note that only the three independent entries in the symmetric $2\times 2$ matrix in the lower right corner has to be specified at each level of the hierarchy. At the level of  wave functions, this amounts to appending new Jastrow factors involving the electrons in the new layer, while leaving intact the ones containing the parent electrons only.
At the operator level, the hierarchal construction is implemented by expressing  the $I+1$'st electron operator in terms of  the $I$'th by	
	\alr{
		V_{I+1}=[\mathcal O V_I]_{r} \ .
		\label{fuse}
	}	
	Here $[\dots ]_r$ denotes a regularization which preserves the topological properties, and $\mathcal O$ satisfies
	\al{
		\mathcal O(\mathbf x_1)\mathcal O(\mathbf x_2)&=(z_1-z_2)^{p_I}\no\mathcal O(\mathbf x_1)\mathcal O(\mathbf x_2)\no\\
		 \mathcal O(\mathbf x_1)V_I(\mathbf x_2)&= (z_1-z_2)^{-1}\no\mathcal O(\mathbf x_1)V_I(\mathbf x_2)\no\\
		  \mathcal O (\mathbf x_1)V_J(\mathbf x_2)&= \noeq\mathcal O (\mathbf x_1)V_J(\mathbf x_2) \noeqr \text{ for } J<I \ ,
	}
	in case the $I$'th condensate is a quasiparticle condensate. For condensates of quasiholes $z_1$ and $z_2$ are replaced by $\bar z_1$ and $\bar z_2$, respectively. The regularization is in no way unique, and the general  form of the hierarchal electron operators is
		\al{
		V_I(\mathbf x)&=\noeq  f_I(\{\pd^{d} \varphi^{k}\})\bar f_I(\{\bar \pd^{d} \bar\varphi^{k}\}) e^{iQ_{IJ}\varphi_L^J+i\bar Q_{IJ}\varphi_R^J}\no (\mathbf x) \ ,
	}
	The functions $f_I$ and $\bar f_I$ are polynomials that are homogeneous in the number of holomorphic and  anti-holomorphic derivatives, respectively, with degree $\sigma_I$ and $\bar \sigma_I$, respectively. Note that, in this realization of the hierarchy, terms with extra factors  $\partial\bar\partial$  of derivatives  can not be added to the wave function albeit having the correct spin according to \eqref{spinvalue}.
	
Although there is a lot of freedom in choosing the functions  $\{f_I\}$ and $\{\bar f_I\}$, there are  restrictions. First,  there should be no poles in the electronic wave functions, and second, the associated correlators should not vanish when convoluted with the anti-symmetrized coherent state kernel. On the plane the first condition is easily implemented by moving all derivatives to the left in the correlation functions, but on the sphere this give chiral wave functions that vanish under convolution with the coherent state kernel, so, \ie, $f_I(\{\pd^{d} \varphi^{k}\})\propto Q_{IJ}\pd^{\sigma_I}\varphi_L^J$ is not allowed (see appendix \ref{app:projection}).

	To make the above discussion more concrete, we consider  the $\nu=(p_0+1/2)^{-1}$ state, \ie, the densest hole condensate on top of the Laughlin $\nu=1/{p_0}$ state. Decomposing the K-matrix as
\begin{align*}
	{\mathbf K} =  \boldsymbol\kappa - \bar{\boldsymbol\kappa} = 		\begin{pmatrix}
		p_0+1& p_0+1\\
		p_0+1&p_0+1
	\end{pmatrix}-
	\begin{pmatrix}
		1& 0\\
		0&1
	\end{pmatrix} \ ,
\end{align*}	
 we get the following two electron operators:
	 \al{
	 	V_1&=\noeq  e^{i\sqrt{p_0+1}\phi_R+i\varphi^1_L}\no \\
		V_2&=[\noeq e^{-i\varphi^1_L+i\varphi^2_L}\nod  e^{i\sqrt{p_0+1}\phi_R+i\varphi^1_L}\noeqr]_{r} \ .
	 }  
	The regularization must not alter the topological properties, and this can be accomplished  \eg by point-splitting and removing the singular part. 
	The point-splitting can however be done  in different ways, as for example, 
		 \mlr{
	 	V_2(\mathbf x)=\lim_{\mathbf x_2\rightarrow \mathbf x}\lim_{\mathbf x_1\rightarrow\mathbf x_2}\no e^{i\sqrt{p_0+1}\phi_R(\mathbf x)+i\varphi^1_L(\mathbf x_1)}\no\times\\
		 \times \no e^{-i\varphi^1_L(\mathbf x_2)+i\varphi^2_L(\mathbf x_2)}\no-\frac1{\bar z_1-\bar z_2}\no e^{i\sqrt{p_0+1}\phi_R(\mathbf x)+i\varphi^2_L(\mathbf x_1)}\noeqr=\\
		=\noeq  \p{i\bar\pd\varphi^1_L-i\bar\pd\varphi^2_L}e^{i\sqrt{p_0+1}\phi_R+i\varphi^2_L}\no (\mathbf x) 
		\label{regularization}
	 }
	 or
	 \mlr{
	 	\tilde V_2(\mathbf x)=\lim_{\mathbf x_2\rightarrow \mathbf x}\lim_{\mathbf x_1\rightarrow\mathbf x_2}\no e^{i\sqrt{p_0+1}\phi_R(\mathbf x)+i\varphi^1_L(\mathbf x_1)}\no \times\\
		\times\no e^{-i\varphi^1_L(\mathbf x_2)+i\varphi^2_L(\mathbf x_2)}\no-\frac1{\bar z_1-\bar z_2}\no e^{i\sqrt{p_0+1}\phi_R(\mathbf x)+i\varphi^2_L(\mathbf x_2)}\no\\
		=\noeq  i\bar\pd\varphi^1_L e^{i\sqrt{p_0+1}\phi_R+i\varphi^2_L}\no (\mathbf x) \ .  \label{ggeneral}
	 }
Linear combinations of the above, or other conventions, are of course also possible. 
	 In this particular case, there is no ambiguity since any term proportional to $\bar\pd\varphi_L^1$ would create poles in the wave function. We are thus forced to take the linear combination where the $\bar\pd\varphi_L^1$ contribution is cancelled, \ie,
	 \alr{
	 	V_2(\mathbf x)=\noeq  i\bar\pd\varphi^2_L e^{i\sqrt{p_0+1}\phi_R+i\varphi^2_L}\no (\mathbf x) \ .
		\label{derchoice}
	 }
	 We could, however, have chosen to represent the operators with more than three bosons, and then there would have been a remaining ambiguity  related to how the Jastrow factors and the derivatives are ordered. The simplest prescription is to move all derivatives to the left, but, as we shall see later, this is not allowed on the sphere since the corresponding wave function will vanish by convolution with the coherent state kernel. Another appealing convention is to factor out the expression $\prod_{i<j} (z_i - z_j)^{1/\nu}$, such that that the remaining part of the correlator is neutral without any background charges\cite{read09}. However this might not work in general, since it could give rise to poles in the wave function, due to the derivatives that act on broken exponents. We now give a convention for the electron operators  at an arbitrary level, that is guaranteed to give regular and non-vanishing wave functions,  and is a generalization of \eqref{derchoice}.
	  To do this, we note that the reason why the choice  \eqref{derchoice} gives a regular wave function is that  the derivatives will only act on a single Jastrow factor. We have constructed operators that achieves this for a general state in the hierarchy. The general expression, which is given in Appendix \ref{operators}, is not very illuminating, so here we just record the special case which is applicable to any state that is obtained by condensation of either only quasiparticles or only quasiholes. This clearly includes the important cases of the positive and negative Jain series. The relevant operators are
	  \ml{
	V_k= \noeq D_{\gamma_k}^{k-1}[\varphi^k] \times\\
	\times e^{i\sqrt{p_0-\gamma_1}\phi^1_R+i\sqrt{p_1-2}\phi^2_{\gamma_2}+\dotsb +i\sqrt{p_{k-1}-2}\phi^k_{\gamma_k}+i\varphi^k_{\gamma_k}}\noeqr \ ,
}
	where $D_{\gamma}^n[f]$ is a polynomial of $f$-derivatives of different order, defined by
	\alr{
		D_{\gamma}^{n}[f]=e^{-if}\pd_{\gamma}^n e^{if} \ . 
		\label{DD}
	}
	The index $\gamma$ labels the chirality as $\phi_+\equiv\phi_R$,  $\phi_-\equiv\phi_L$, $\pd_+\equiv \pd$ and $\pd_-\equiv \bar\pd$. \label {general}
	
	Using the techniques developed in  Refs.  \onlinecite{suorsa11a, suorsa11b} closed form expressions are not possible for the operators in an arbitrary mixed state, \ie, a state formed by condensates of both quasiparticles and quasiholes, while here we give such  a general expression (see Appendix \ref{operators}).
	The reason can be traced back to an assumption in Refs. \onlinecite{suorsa11a, suorsa11b}, where a two fluid picture is employed, each containing particles with the same charge. Defining $\boldsymbol\tau=\mathbf Q \mathbf c$ and $\bar{\boldsymbol\tau}=\bar{\mathbf Q} \bar{\mathbf c}$ this condition amounts to taking	 
	 $\boldsymbol\tau\propto\bar{\boldsymbol\tau}\propto \mathbf t$. This constraint does not follow from any physical principle, and it gives a complicated condition on the allowed decompositions of $\mathbf K$ into $\boldsymbol \kappa$ and $\bar{\boldsymbol\kappa}$. In fact, it must  be solved on a case by case basis, and it is not clear that a solution always exists. 
	
	By relaxing the requirement that all charges in the two fluids should be the same, we can substantially simplify the representative wave functions for the mixed states and give  explicit expressions for all the wave functions,   rather than have to construct them case by case. The earlier proposed wave functions for the mixed states were also somewhat unsatisfactory in that they contained Jastrow factors to high powers and thus were hard to compute numerically. 
	Using the method proposed  in this paper, there is no such difference in complexity between the mixed states and those obtained from condensing excitations of only one type.
		
	\section{The spherical geometry}
	\subsection{Choosing coordinates and gauge}
	When working with a conformal field theory it is most convenient to use isothermal coordinates, and from now on $\mathbf x=(x,y)$ will denote the isothermal coordinates of a point on the sphere and $z$ will denote the complex coordinate $z=x+i y$. In isothermal coordinates the metric is a Kronecker delta up to a scale factor,
	\al{
		g_{\mu\nu}&= 2 R e^{2\omega(\mathbf x)} \delta_{\mu\nu}&
		\omega(\mathbf x)&=-\ln(1+|\mathbf x|^2)\ .
	} 
	Here $R$ is the radius of the sphere, which we for notational simplicity from now will set to one half, \ie, $R\equiv1/2$.
	
	For a homogeneous magnetic field the magnetic flux density is proportional to the surface element, and with our conventions, with $2R\equiv1$ and $\phi_0/2\pi=\hbar c/e\equiv1$, the proportionality constant will be twice the number of flux quanta,
\al{
	\mathcal B= 2 N_\phi\, \sqrt g \,\dd x^1\wedge \dd x^2\ .
}
On the sphere, the total magnetic field is not arbitrary but has to fulfill the Dirac quantization condition,
\al{
	\int \mathcal B&=2\pi n \quad n\in \Z\  &&\Leftrightarrow& N_\phi&\in \Z \ .
}	
	To get explicit expressions for the wave functions we must choose a gauge. With the isothermal coordinates it is simplest to work with a gauge potential which is well defined in a arbitrarily large open region around $\mathbf x=\mathbf 0$, and rotationally symmetric around $\mathbf x=\mathbf 0$. This implies the Dirac gauge\cite{dirac31}
\alr{
	\mathcal A=i \frac {N_\phi}2\frac{z \dd\bar z-\bar z \dd z}{1+z \bar z}\ .
	\label{gauge}
}
	Notice that this gauge is well defined everywhere except at $z\rightarrow \infty$.
	\subsection{Conformal transformations and the massless bosons}
	The transformations of CFT operators on the sphere differ from those on the plane because of the conformal factor in the metric. Under a general conformal transformation
\al{
	z\rightarrow z'=\frac{a z+b}{c z +d}\quad \text{ with } ad-bc\neq 0 \ ,
}
a quasi-primary field $\mathcal O(\mathbf x)$ of weight $h,\bar h$ transforms as
\alr{
	\mathcal O(\mathbf x)=\p{\frac{\pd z'}{\pd z}}^h \p{\frac{\pd \bar z'}{\pd \bar z}}^{\bar h}e^{(h+\bar h)(\omega(z')-\omega(z))}\mathcal O(\mathbf x') \ .
	\label{quasiprimary}
}
Notice the extra metric dependent factor which does not appear in a planar geometry. 

As on the plane, we will construct all fields in terms of massless bosons. The Greens function of the Laplacian, and therefore the two-point-function of the bosons, is 
	\alr{
		\braket{\varphi(\mathbf x_1)\varphi(\mathbf x_2)}=-\ln|\mathbf x_1-\mathbf x_2|^2-\omega(\mathbf x_1)-\omega(\mathbf x_2)
		\label{greensfunction}
	}
	up to an unimportant additive constant. As on the plane, there are in principle many different ways we can split our bosons into chiral and anti-chiral pairs, which results in wave functions in different gauges. The simplest choice is to again use the operator $L_3=z\pd-\bar z\bar\pd$ which will amount to wave functions in the Dirac gauge \eqref{gauge}.
	
	On the sphere, the Laplacian is, up to a multiplicative constant, the angular momentum operator $\mathbf L^2$. The space with a definite eigenvalue of the Laplacian is, thus, finite-dimensional, and each space contain one function which is rotation symmetric around $\mathbf x=\mathbf 0$. These rotation invariant states therefore give a non-zero contribution to the Greens function, \ie, the $\omega$-terms in \eqref{greensfunction}. Because of the contribution of these modes, we have to take special care when defining $P_L$ and $P_R$. 
	As we will see, the electron operators must have well defined conformal weights to describe quantum Hall states. From the transformation property of quasi-primary fields \eqref{quasiprimary} we see that a chiral vertex operator should have the same $\omega$ dependence as an anti-chiral, and half compared to the full field. We can thus conclude that if the chiral and anti-chiral vertex operators are to have well defined conformal weights they must get the same contribution from the rotation invariant parts, and the sum of the parts must add up to the full Greens function. We therefore define $P_R$ as the identity on all functions except negative eigenstates of $L_3=z\pd-\bar z\bar\pd$, which get annihilated, and zero eigenstates of $L_3$, which all except for the zero mode of the Laplacian get multiplied by one half. $P_L$ is defined in the same way except that it annihilates positive eigenstates instead of negative. The two-point function of chiral operators is therefore
	\al{
		\braket{\varphi_R(\mathbf x_1)\varphi_R(\mathbf x_2)}=-\ln(z_1-z_2)-\frac12\omega(\mathbf x_1)-\frac12\omega(\mathbf x_2) \ .
	}

\section{Rotation invariance, the background operator, and the neutrality condition}
\label{rotation-invariance}
The quantum Hall wave functions must realize the symmetry of the microscopic Hamiltonian up to a possible gauge transformation. Thus, under a rotation $\mathbf x\rightarrow \mathbf x^\prime$ we have:
\al{
	\mathcal A=A_i(x) \dd x^i&\rightarrow A_i(x^\prime) \dd x^{\prime i}\\
	 \Psi(\{x_i\})&\rightarrow e^{-i\sum_i \Lambda(x^\prime_i)} \Psi(\{x^\prime_i\}) \ ,
}
with $\dd \Lambda=A_i(x) \dd x^i-A_i(x^\prime) \dd x^{\prime i}$. This transformation property we call magnetic rotation invariance, in analogy with magnetic translation invariance on the plane. Stated differently, a magnetically rotation invariant wave function satisfies
\alr{
	\Psi(\{x_i\})&=e^{-i\sum_i \Lambda(x^\prime_i)} \Psi(\{x^\prime_i\})
	\label{lambdaphase}
}
under the rotation $\mathbf x\rightarrow \mathbf x'$, which in our coordinates amounts to
\alr{
	z \rightarrow z^\prime&=\frac{u z-v^*}{v z+u^*} &\text{wi}&\text{th}& uu^*+vv^*&\neq0 \ .
	\label{rotation}
}
Inserting \eqref{rotation} in \eqref{gauge} we get 
\al{
	\dd\Lambda= i \frac{N_\phi}2\, \dd\p{\ln\p{\frac{v^* \bar z+u}{vz +u^*}}}\ .
}
Thus, under the rotation \eqref{rotation} a wave function for a rotation invariant state acquire the phase,
\alr{
	\Psi(\{x_i\})=\prod_i\p{\frac{v^* \bar z_i+u}{vz_i +u^*}}^{-N_\phi/2}\Psi(\{x^\prime_i\}) \ .
	\label{wftransformation}
}

We now compare the transformation of magnetically rotation invariant wave functions \eqref{wftransformation}, to the transformation of quasi-primary fields \eqref{quasiprimary}. The rotations of the sphere are isometries, so under the transformation \eqref{rotation} we have
\al{
	e^{2(\omega(z^\prime)-\omega(z))}=\frac{\pd z}{\pd z^\prime}\frac{\pd \bar z}{\pd \bar z^\prime} \ .
}
Consequently, a quasi-primary field $\mathcal O(\mathbf x)$ with conformal weights $h,\bar h$, transforms as 
\alr{
	\mathcal O(\mathbf x)=\p{\frac{\pd_z z^\prime}{\pd_{\bar z}\bar z^\prime}}^{\frac{h-\bar h}2}\mathcal O(\mathbf x^\prime) =\p{\frac{v^* \bar z+u}{vz +u^*}}^{h-\bar h}\mathcal O(\mathbf x^\prime) \ .
	\label{invaraince}
}
From \eqref{invaraince} we see that wave functions constructed as correlators of operators with spin $h-\bar h=-N_\phi/2$ will be magnetic rotation invariant\footnote{One can of course object and say that it is the position basis wave functions and not the coherent state wave functions that should transform in this way. But the integration with the coherent state kernel \eqref{kernel} is equivalent with LLL projection which can be written as 
\al{
	P_{LLL}= \prod_{l=N_\phi/2+1}^\infty \frac{l(l+1)-\mathbf L^2}{l(l+1)-\frac{1}8N_\phi(N_\phi+4)} \ .
}
Here $\mathbf L^2=\sum_i L^2_i$ where $\{L_i\}_{i=1,2,3}$ span the Lie algebra af the magnetic rotations and obey the usual angular momentum algebra. So we can conclude that if a coherent state wave function is magnetic rotation invariant, \ie annihilated by $L_i$ for $i=1,2,3$, so is the position basis wave function. }.

	For consistency, we also have to add the spin curvature contribution to the background. On the sphere there is a shift $\mathcal S$ in the relation between magnetic flux quanta and number of particles\cite{haldane83},  $N=\nu (N_\phi+\mathcal S)$.  This comes about since moving a particle with spin on a curved surface is equivalent to moving a charged particle in a magnetic field, with the charge replaced by spin and the magnetic field replaced by the Ricci curvature $\mathcal R$. So we should add a term of the type $\frac{-i}{4\pi}\int d^2x\sqrt g \mathcal R \Sigma_I \varphi^I$, and a similar one for the anti-chiral part, to the background term in the action. 	
		The spin of an operator can be read off from the response to a rigid rotation, \ie $z\rightarrow z e^{i\theta}$ and $\bar z e^{-i\theta}$, and an electron operator transforms as
	$$
		V_I(\mathbf x)\rightarrow e^{-is_I\theta+it_I\frac{N_\phi}2\theta}V_I(\mathbf x) \ .
	$$
	The second, extensive, term arises from the background\footnote{The background term in the action changes the conformal weight for the vertex operators $h_\alpha \rightarrow h-Q \tau_\alpha$ and $\bar h_\alpha \rightarrow \bar h-Q \bar\tau_\alpha$, see for example chapter nine in Ref. \onlinecite{francesco97}. In appendix \ref{transformation} we explicitly check that the operators have the correct transformations properties.} while the first term arises from the rotation of the electron it self. We have seen that if the correlators of the electron operators are to be rotation invariant the electron operators need to have spin $-Q$, see equation \eqref{invaraince}. Therefore the first factor $e^{-i s_I \theta}$ should be cancelled by the contribution from the curvature in the background. This results in the constraint
	\alr{
		\mathbf Q \boldsymbol \Sigma-\bar{\mathbf Q} \boldsymbol {\bar\Sigma}=\mathbf s \ ,
		\label{constraint3}
	}
	where $\mathbf s$ is the spin vector in the multilayer representation. The extra term in the background operator also modifies the neutrality conditions, which become,
	\se{neutralitycondition}{
		\sum_I N_I Q_{IJ}&=N_\phi c_J + 2\Sigma_J \\
		\sum_I N_I \bar Q_{IJ}&=\bar N_\phi \bar c_J + 2\bar \Sigma_J \ .
	}
	The neutrality condition and the constraint \eqref{constraint3} determines the spin vectors to be
	\al{
		\boldsymbol\Sigma&=\mathbf Q^T \mathbf K^{-1}\mathbf s &\text{an}&\text{d}& \boldsymbol{\bar\Sigma}&=\mathbf {\bar Q}^T \mathbf K^{-1}\mathbf s \ .
	}
	Furthermore, the neutrality conditions \eqref{neutralitycondition} imply that
	\al{
		\sum_I N_I K_{IJ}&=N_\phi+2s_J \ ,
	}
	which in turn implies that the shift is
	\al{
		\mathcal S=\frac{2}{\nu}\sum_{IJ} K^{-1}_{IJ}s_J \ .
	}
	This formula, which is in  agreement with the general result given by Read\cite{read09}, says that the shift is twice the average conformal spin of the fields representing the electrons with no background term present.

\section{The explicit form of the operators and the wave functions} 
We want to define quantities analogous to the derivatives in the wave functions in planar geometry that originate from operators like \eqref{fuse}. On the sphere we have to be more careful with the regularization. We need to make sure that the resulting wave functions are square integrable. The simplest way to do this is to define the regularization by point-splitting only in the coordinate $|z|$ but not in $\arg(z)$. As an example we do the calculation \eqref{regularization} again but now on the sphere. We start by considering the electron operator and the hole operator at two different points and take $\arg(z)=\arg(z_1)=\arg(z_2)$ for the reason just mentioned.
	\ml{
	\no e^{i\sqrt{p_0+1}\phi_R(\mathbf x)+i\varphi^1_L(\mathbf x_1)}\nod e^{-i\varphi^1_L(\mathbf x_2)+i\varphi^2_L(\mathbf x_2)}\noeqr =\\
	\frac{\prod_{i=1}^2(1+z_i\bar z_i)^{1/2}}{\bar z_1-\bar z_2}\no e^{i\sqrt{p_0+1}\phi_R(\mathbf x)+i\varphi^1_L(\mathbf x_1)-i\varphi^1_L(\mathbf x_2)+i\varphi^2_L(\mathbf x_2)}\no 
	}
	Taylor expanding the last factor gives
	\ml{
		\no e^{i\sqrt{p_0+1}\phi_R(\mathbf x)+i\varphi^1_L(\mathbf x_1)-i\varphi^1_L(\mathbf x_2)+i\varphi^2_L(\mathbf x_2)}\noeqr =\noeq \bigl[1+\\
		(\bar z_2-\bar z_1)\p{i\bar\pd\varphi^1_L-i\bar\pd\varphi^2_L}+(z_2-z_1)\p{i\pd\varphi^1_L-i\pd\varphi^2_L}\bigr]\times \\
		\times e^{i\sqrt{p_0+1}\phi_R(\mathbf x)+i\varphi^2_L(\mathbf x_1)}\no+\mathcal O(|\mathbf x_1-\mathbf x_2|^2) \ .
	}
	Hence subtracting 
	\al{
	\frac{(1+z_1\bar z_1)^{1/2}(1+z_2\bar z_2)^{1/2}}{\bar z_1-\bar z_2}\no e^{i\sqrt{p_0+1}\phi_R(\mathbf x)+i\varphi^2_L(\mathbf x_1)}\no
	}
	 from 
	 \al{
	 \no e^{i\sqrt{p_0+1}\phi_R(\mathbf x)+i\varphi^1_L(\mathbf x_1)}\nod e^{-i\varphi^1_L(\mathbf x_2)+i\varphi^2_L(\mathbf x_2)}\no
	}
	 and taking the limit $\lim_{\mathbf x_2\rightarrow \mathbf x}\lim_{x_1\rightarrow  x_2}$, we get
	\ml{
	(1+z\bar z)\no\p{\p{i\bar\pd\varphi^2_L-i\bar\pd\varphi^1_L}+\frac z{\bar z}\p{i\pd\varphi^2_L+i\pd\varphi^1_L}}\times\\
	\times e^{i\sqrt{p_0+1}\phi_R+i\varphi^2_L}\no(\mathbf x) \ .
	}
The holomorphic derivative $\frac z{\bar z}\pd$ will only act on the zero-modes of $L_3$ in $\varphi_L^2$ and double their contribution as compared to just having the anti-holomorphic derivative. The term with the anti-holomorphic derivative can therefore be cancelled by using the factor $\bar \pd \varphi^2$ which contains the full field, not just the anti-chiral $\bar \pd\varphi^2_L$. As discussed in section \ref{convention}, we need to make the $\bar\pd \varphi^1$ piece vanish to avoid poles in the electron coordinates. For that reason we choose the linear combination 
	\al{
		V_2(\mathbf x)=(1+z\bar z)D_{-}^1[\varphi^2]\no e^{i\sqrt{p_0+1}\phi_R(\mathbf x)-i\tilde\varphi^2_R(\mathbf x)}\bar \pd e^{i\varphi^2(\mathbf x)}\no\ \, ,
	}
	of pointsplittings at different positions.  It is now important that the derivative act on the full operator, $\varphi_2$, not only the anti-chiral part. Besides the double counting of the zero eigenstates of $L_3$ the logic is exactly the same as on the plane, and we can use exactly the same operators if we just add the metric dependence to $D$, see \eqref{DD}. That  is, we redefine $D$ as
	$$
		D_{\gamma}^n [f(\mathbf x)]=e^{-n\omega(x)} e^{-if(\mathbf x)}\pd_{\gamma}^n e^{if(\mathbf x)} \ .
	$$

	 As an example we can calculate the wave function for the observed mixed state at $\nu=5/7$ which is a maximally dense particle condensate on top of the $\nu=2/3$ state. The electron operators are (see appendix \ref{operators})	
	\al{
	 	V_1&=\noeq e^{i\phi^1_L+i\varphi_L^1}\no \\
	 	V_2&=D^1_-[\varphi^2]\noeq e^{i\phi^1_L+i\varphi^2_L+i\psi^2_R+i\chi^2_R}\no\\
	 	V_3&=D^1_+[\psi^3]\noeq e^{i\phi^1_L+i\sqrt{2}\phi^2_R+i\varphi^2_L+i\psi^3_R+i\chi^2_R}\no
	}
	and the coherent state wave function $\Braket{V_1(\mathbf x_1)\dotsb V_3(\mathbf x_{N})}$ will consist of three factors: one from the $L_3=0$ parts of the Greens function,
	\ml{
		\prod_{i\in S_1}(1+z_i\bar z_i)^{-N_\phi/2(\tau_1+\bar\tau_1)}\times\prod_{i\in S_2}(1+z_i\bar z_i)^{-N_\phi/2(\tau_2+\bar\tau_2)}\times\\
		\times\prod_{i\in S_3}(1+z_i\bar z_i)^{-N_\phi/2(\tau_3+\bar\tau_3)}, 
	}
	one from the chiral part,
	\al{
		\prod_{i<j}(z_i-z_j)^2 \times(2-2) \prod_{i\in S_3}\p{\pd-\frac{(N_3-1)\bar z_i}{1+z_i\bar z_i}}\times(3-3) \ ,
	}
	and one from the anti-chiral part
	\ml{
		(\bar 1-\bar 1)(\bar 2-\bar 2)(\bar 3-\bar 3)^2(\bar 2-\bar 3)^2\times\\
		\times \prod_{i\in S_2}\p{\bar \pd-\frac{(N_2-1) z_i}{1+z_i\bar z_i}}\times(\bar 2-\bar 2) \ .
	}
	Here we used the same short hand notation as in the introduction, with $S_1=\{1,\dotsc N_1\}$  and with $S_2$ and $S_3$ analogously defined. To get the wave functions in position basis we must  convolute with the coherent state kernel. Although a bit more involved this can just as on the plane be done algebraically. We show this in appendix \ref{app:projection}.	
\section{Excitations}
To construct the quasiholes one do exactly as on the plane, namely by inserting a quasihole operator
\al{
	H_{\mathbf l}(\mathbf x_h)=   \noeq e^{-i(q_{\mathbf l})_I\varphi^I_R-i(\bar q_{\mathbf l})_I\bar \varphi^I_L}\noeqr (\mathbf x_h) \ ,
}
in the correlator. It is instructive to look at the spin of the operator, 
\al{
	h-\bar h=\frac{\mathbf l \mathbf K^{-1} \mathbf t N_\phi}2+\mathbf l \mathbf K^{-1}\mathbf s+\frac12\mathbf l \mathbf K^{-1}\mathbf l \ ,
}
see appendix \ref{transformation}. Appart from the part from the magnetic field, there is an additional part $\mathbf l \mathbf K^{-1}\mathbf s+\frac12\mathbf l \mathbf K^{-1}\mathbf l$ which implies that the particle carries an (orbital) spin. This value coincides with the one predicted by the CS theory in Ref. \onlinecite{wen95}. 

The quasiparticle operators introduced in Ref. \onlinecite{hansson09a, *hansson09b} can, just  as the quasihole operators, be constructed on the sphere. 
The details, which includes using a spherical rather than planar coherent state kernel when fusing an inverse quasihole operator with an electron operator, are fairly straightforward to work out and will not be reported here.

\section{Summary and Outlook}
	In this paper we have generalized the representative wave functions, developed in Refs. \onlinecite{hansson07}-\onlinecite{suorsa11b}, to  spherical geometry. 
	We explicitly constructed all states in the hierarchy, determined the shifts, and showed that for states in the Jain series, our wave functions are identical to those obtained using composite fermions. 	 	
	The wave functions for mixed states proposed here  are better suited for numerical study than those previously proposed. This is both because the sphere is the geometry of choice for  numerics, and because the wave functions themselves are simpler, as discussed in detail above. It should be possible to make comparison with  exact diagonalization studies for small systems and thus get some insight into which mixed states might be most easily realized with realistic potentials. Such studies have been very successful for the composite fermion states, and our methods can hopefully extend those studies to a larger part of the Haldane-Halperin hierarchy. 
	
\begin{acknowledgments}
I thank Hans Hansson for suggesting this project and for many valuable discussions and for helpful comments on the manuscript. 
I also thank Maria Hermanns for many helpful discussions, and Eddy Ardonne, Sören Holst and Mikael Fremling for reading and commenting on the  manuscript.
\end{acknowledgments}\appendix
\section{The Landau problem on the sphere}
The  Landau problem, {\em \ie} a charged particle moving in a homogenous magnetic field, on the sphere, was originally solved  by Igor Tamm in 1931\cite{tamm31}, and has been the subject of many subsequent papers. To make the comparison between the wave functions developed here, and the CF wave functions we need the single particle wave functions in a slightly different form than the conventional. The easiest way to show that they take this form is to solve the one particle problem again.

In index notation the Hamiltonian reads
$$
	H=\frac{1}{\sqrt g}\p{-i\pd_\mu-\mathcal A_\mu}\sqrt g g^{\mu\nu}\p{-i\pd_\nu-\mathcal A_\nu} \ ,
$$
which can be rewritten as
\al{
	H=a^\dagger_Q a_Q +Q=a_{Q-1} a^\dagger_{Q-1} -Q \ ,
}
where $a_Q=i(1+z \bar z)\bar \pd+iQ z$, and $Q\equiv N_\phi/2\text{ sign}(\mathcal B)$. The dagger denotes Hermitian conjugate%
\footnote{Here $\sqrt g\neq \text{const.}$ so Hermitian conjugation picks up a contribution from the measure.}
 , which means that $a^\dagger_Q=i(1+z \bar z) \pd-i(Q+1) \bar z$.
 
If $Q>0$ there are states, with well-defined norm, annihilated by $a_Q$ and if $Q<0$ there are states annihilated by $a^\dagger_{Q-1}$. Since   $a^\dagger_Q a_Q$ is positive definite we see that these states span the LLL. For notational simplicity we assume $Q≥0$. The case $Q≤0$ is completely analogous, just that $z$ and $\bar z$ change place, $-Q$ and $Q$ change place, and $m$ and $-m$ change place. 
 
The Hamiltonian commutes with the generators of magnetic rotation, and the generator of magnetic rotation around $\mathbf x=\mathbf 0$ is $L_3=z\pd-\bar z\bar\pd-Q$. In this section $L_3$ will denote the generator of magnetic rotation around $\mathbf x=\mathbf 0$ rather than ordinary rotation as previously. We can therefore use the eigenfunctions of $L_3$ as a basis for the LLL. The normalized eigenfunctions $\Psi_{Q0m}$ with LL index $0$ and $L_3$ eigenvalue $m$ read
\al{
	\Psi_{Q0m}=\sqrt{2\pi (2Q+1)\binom{2Q}{Q-m}}z^{m+Q}(1+z \bar z)^{-Q} \ .
} 
Notice that the requirement that the wave functions are normalizable imply $-Q≤m≤Q$. Using 
\al{
	a_Q a^\dagger_R - a^\dagger_{R+1} a_{Q+1}&=(Q+R+2)\\
	a^\dagger_R L_3^Q-a^\dagger_R L_3^{Q+1}&=0 
}
we get
\al{
	\sqrt{E_{Q\,n+1}+Q+1}\Psi_{Q\, n+1\, m}&=a^\dagger_Q \Psi_{Q+1\,n\,m}\\
	E_{Q\,n+1}&=E_{Q+1\,n}+2Q+1
}
which used repeatedly reveals
$$
		E_{Q\, n}=(Q+n)(Q+n+1)-Q^2 \ ,
$$
and
\mlr{
	\Psi_{Q\, n\, m}=\sqrt{2\pi (2Q+1)\binom{2Q}{Q-m}}\times\\
	\times\p{\prod_{p=0}^{n-1} \frac{a_{Q+p}^\dagger}{\sqrt{n(2Q+n-1)+p+1}}}\times\\
	\times z^{Q+n+m}(1+z\bar z)^{-Q-n}
	\label{singleparticlefunction}\ .
}
The spherical coherent state wave function $\braket{z|\xi}$ can, as on the plane, be viewed as the LLL projection of the Dirac delta function,
\ml{
	\braket{z|\xi}= \sum^{Q}_{m=-Q} \Psi_{Q0 m}(z,\bar{z}) \Psi_{Q0m}^*(\xi,\bar{\xi})=\\
	=\frac{(2Q+1)(1+z\bar\xi)^{2Q}}{ (1+z\bar{z})^{Q}(1+\xi\bar{\xi})^{Q}} \ .
}
The many body coherent state kernel is therefore
\mlr{
		\braket{z_1,\dotsc, z_N|\xi_1,\dotsc, \xi_N}\propto\\
		\propto\mathcal A \prod_i \frac{(2Q+1)(1+z_i\bar\xi_i)^{2Q}}{ (1+z_i\bar z_i)^{Q}(1+\xi_i\bar\xi_i)^Q} \ .
		\label{kernel}
}
\section{Equivalence with composite fermion wave functions}
\label{app:CF}
To simplify notation we will only look at the positive Jain series. The treatment of the reverse flux attachment Jain series is however completely analogous, the only difference is that $z_i$ and $\bar z_i$ change place in the filled Landau levels, in the composite fermion language. 

Using the conventions developed in this article the coherent state wave functions for the Jain series are given by the correlator
\al{
	\Braket{\prod_{i\in S_1} V_1(\mathbf x_i)\times\dotsb\times\prod_{i\in S_n} V_n(\mathbf x_i)} \ ,
}
where $V_k(\mathbf x)=D_+^{k-1}[\varphi^k]\no  e^{i\sqrt{p_0-1}\phi^1_R+i\varphi^k_R}\no(\mathbf x)$ and $N_k=|S_k|$ is by the neutrality condition equal to $N_k=\frac1nN+2k+1-n$. Written out the correlator is
	\mlr{
		\prod_i(1+z_i\bar z_i)^{-Q}\times\prod_{i<j}(z_i-z_j)^{2p}\times(1-1)\prod_{i\in S_2} \Gamma_{N_2}(\mathbf x_i)\times\\
		\times(2-2)\times\dotsb\times\prod_{i\in S_n} \Gamma_{N_n}^{n-1}(\mathbf x_i)(n-n)\ ,
		\label{CFT:composite}
	}
	where $\Gamma_A=\pd-\frac {\bar z A}{(1+z\bar z)}$.
	
The composite fermion wave function\footnote{The development of the composite fermion  wave functions and the investigation of them is work that is presented in several articles. The main contributor to this work is J. K. Jain and references to most of the original work on composite fermions can be found in his book\cite{jain07}. }\nocite{jain07} at filling fraction $\nu=\frac n{2pn+1}$ with $N=n k$ number of particles is defined as
	\ml{
		\Psi^{CF}_{\nu=\frac n{2pn+1}}(\{z_i,\bar z_i\})\propto P_{LLL}\prod_i(1+z_i\bar z_i)^{p-npk}\times\\
		\times\prod_{i<j}(z_i-z_j)^{2p} \Phi_n^{k-n}(\{z_i,\bar z_i\}) \ ,
	}
	where $P_{LLL}$ is the projection operator onto the lowest Landau level, and $\Phi_n^{k-n}$ is $n$ filled Landau levels at flux $k-n$. 
	 Using \eqref{singleparticlefunction} and the Vandermond identity we see that the $l$'th LL at flux $k-n$ ($\Psi_l^{k-n}(\{\mathbf x_i\})$) can be written as 
	\ml{
		\Psi_l^{k-n}(\{\mathbf x_i\})\propto\prod_i\prod_{m=0}^{l-1}a_{(k-n)/2+m}^\dagger(\mathbf x_i)\times\\
		\times\prod_i (1+z_i\bar z_i)^{(n-k)/2-l}\prod_{i<j}(z_i-z_j) \ .
	}
	Taking all factors $(1+z_i\bar z_i)$ from the $a^\dagger$'s and commute them through, to the furthest left, we turn the $a^\dagger$'s into $\Gamma_{N_l}$'s, and we get
	\ml{
		\Psi_l^{k-n}(\{\mathbf x_i\})=\propto \prod_i(1+z_i\bar z_i)^{(n-k)/2}\times\\ 
		\times\prod_i \Gamma_N^l(\mathbf x_i)\prod_{i<j}(z_i-z_j) \ .
	}
	 We thus see that we can rewrite the composite fermion wave functions into the form \eqref{CFT:composite}.	

\section{Explicit form of the operators for the mixed states}
\label{operators}
As compared to the states in the hierarchy reached by only quasihole or only quasiparticle condensations the mixed states are a little more complicated because there are both holomorphic and anti-holomorphic derivatives, and in case of a mix condensation all four of the question marks in \eqref{matrix} change, as opposed to only the three in the bottom corner otherwise. As an example we can look at a state which is a particle condensate on top of a second level hole condensate:
	\al{
	 V_1&=\noeq e^{i\sqrt{p_0+1}\phi^1_L+i\varphi_L^1}\no \\
	 V_2&=\noeq D^1_-[\varphi^2]e^{i\sqrt{p_0+1}\phi^1_L+i\sqrt{p_1-2}\phi^2_L+i\varphi^2_L+i\psi^0_R+i\chi_R}\no\\
	 V_3&=\noeq D^1_-[\varphi^2]D^1_+[\psi^3]e^{i\sqrt{p_0+1}\phi^1_L+i\sqrt{p_1-2}\phi^2_L+i\sqrt{p_2+2}\phi^2_R}\times\\
	&\mspace{250mu} \times e^{i\varphi^2_L+i\psi^1_R+i\chi_R}\noeqr \ .
	}
	The general expression for the operators in any state can, of course, be written down using the same conventions. The generality makes the notation a bit involved, but for the sake of completeness we write it down:
	\ml{
	V_k=\noeq D^{k_{\gamma_1}}_{\gamma_1}[\varphi^{k_\varphi}_{\gamma_{k_\varphi}}]D_{-\gamma_1}^{k_{\gamma_1}}[\psi^{k_\psi}_{\gamma_{k_\psi}}]	\times\\
	 \exp\Bigl(i\sqrt{p_0-\gamma_1}\phi^1_R+i\sqrt{p_1-2\gamma_1\gamma_2}\phi^2_{\gamma_2}+\dotsb\\
	+i\sqrt{p_{k-1}-2\gamma_{k-1}\gamma_k}\phi^k_{\gamma_k}+i\varphi^{k_{\gamma_1}+1}_{\gamma_1}+\delta_k\bigl(i\psi^{k_{-\gamma_1}}_{-\gamma_1}+i\chi_{\gamma_1}\bigr)\Bigr)\noeqr \ .
	}
	Here $k_{\gamma_1}$ denote the number of condensates of particles which have charge with the sign $\gamma_1$ up to level $k$ and equivalent for $k_{-\gamma_1}$.  When the elements in $\{\gamma_i\}_{i=1,\dotsc k}$ all have the same sign $\delta$ yields the value $0$ and $1$ otherwise.

\section{Transformation properties of the operators}
	\label{transformation}

	In this section we will show that in spherical geometry, with the action
	\ml{
		S[\varphi]=-\frac1{8\pi}\int d^2x\sqrt g \p{\sum_i \varphi^i \Delta \varphi^i+\sum_i \phi^i \Delta \phi^i}\\
		-\frac i{2\pi}\int\mathcal B \p{ c_I \varphi^I+\bar c_I \varphi^I}\\
		-\frac i{4\pi}\int d^2 x \sqrt g\, \mathcal R\p{\Sigma_I \varphi^I+{\bar\Sigma}_I\phi^I}
	}
	operators of the form 
	\ml{
		\mathcal O_\alpha(\mathbf x)=\noeq D_+^{\sigma_\alpha}\left[y_I\varphi^I\right]D_-^{\bar\sigma_I}\left[\bar y_I\bar \varphi^I\right]e^{i r_I\varphi^I_R+i\bar r_I\phi^I_L}\no(\mathbf x)
	}
	will, under isometries, transform as a quasi-primary field, with conformal weights 
\se{conformalweight}{
	h_\alpha&=\frac12\p{\mathbf r^T \mathbf r-N_\phi \mathbf r^T\mathbf c}-\mathbf r^T\boldsymbol \Sigma+\sigma_\alpha \\ 
	\bar h_\alpha&=\frac12\p{\bar{\mathbf r}^T \bar{\mathbf r}-N_\phi\bar{ \mathbf r}^T\bar{\mathbf c}}-\bar{\mathbf r}^T\bar{\boldsymbol \Sigma}+\bar \sigma_\alpha \ .
}
The index $\alpha$ denotes the set $\{\mathbf r,\mathbf y,\boldsymbol\sigma, \mathbf {\bar r},\mathbf {\bar y},\boldsymbol{\bar \sigma}\}$. That an operator has conformal weights $(h,\bar h)$ means that correlation functions of these operators have the property
\ml{
	\Braket{V_1(\mathbf x_1)\dotsb V_N(\mathbf x_N)}=\\
	=\p{\frac{\pd z'_1}{\pd z_1}}^h \p{\frac{\pd \bar z'_1}{\pd \bar z_1}}^{\bar h}e^{(h+\bar h)(\omega(z'_1)-\omega(z_1))}\times\dotsb\\
	\dotsb\times\p{\frac{\pd z'_N}{\pd z_N}}^h \p{\frac{\pd \bar z'_N}{\pd \bar z_N}}^{\bar h}e^{(h+\bar h)(\omega(z'_N)-\omega(z_N))}\times\\
	\times\Braket{V_1(\mathbf x_1')\dotsb V_N(\mathbf x_N')} \ .
}
To prove this we first look at the neutrality condition. The correlation functions vanish unless the total coefficient in front of each zero mode vanishes, \ie, 
\alr{
		\sum_{\alpha=1}^N \mathbf r_\alpha&=N_\phi \mathbf c+\boldsymbol 2\Sigma  \label{neutrality1}\\
		\sum_{\alpha=1}^N \mathbf {\bar r}_\alpha&=N_\phi \mathbf {\bar c}+2\boldsymbol {\bar \Sigma} \label{neutrality2}\ .
}
The non-vanishing correlation functions will be 
\ml{
	\Psi(\{\mathbf x_i\})=\Braket{V_1(\mathbf x_1)\dotsb V_N(\mathbf x_N)}\propto\\
		\prod_{\alpha<\beta}(z_\alpha-z_{\beta})^{\mathbf Y\cdot \mathbf Y'}(1+z_{\alpha}\bar z_{\alpha})^{\frac{-\mathbf r\cdot \mathbf r'}2}(1+z_{\beta}\bar z_{\beta})^{\frac{-\mathbf r\cdot \mathbf r'}2}\times\\
		\prod_\alpha(1+z_\alpha\bar z_\alpha)^{\sigma_\alpha}\p{\pd_\alpha+\frac{C\bar z_\alpha}{1+z_\alpha\bar z_\alpha}}^{\sigma_\alpha}\prod_{\alpha<\beta}(z_\alpha-z_{\beta})^{\mathbf y\cdot \mathbf y'}\\
		\times\text{ anti-chiral ,} 
}
where `anti-chiral' denote an equivalent term from the anti-chiral part of the operators and $\mathbf Y\equiv \mathbf r_\alpha-\mathbf y_\alpha$, $\mathbf Y'\equiv \mathbf r_{\beta}-\mathbf y_{\beta}$, $\mathbf y\equiv  \mathbf y_\alpha$, and $\mathbf y\equiv  \mathbf y_{\beta}$. $C$ is here a  constant that depend on $\{\mathbf \alpha\}$. Since its precise value does not matter for the forthcoming discussion we do not specify it. To get this expression we have assumed that $\mathbf y$ is orthogonal to $\mathbf Y$. If this was not the case the correlation function would have been sums of terms of this form, with the derivatives at different positions. As we will see, the exact placement of the derivatives will not matter in the arguments given below. So the assumption $\mathbf y\perp \mathbf Y$ is only for convenience. 
\subsection{Rotation}
We now want to show that the correlation functions $\Psi(\{\mathbf x_i\})$ are invariant under a conformal transformation $\mathbf x\rightarrow \mathbf x'$ together with $\Psi\rightarrow \p{\frac{\pd z'_1}{\pd z_1}}^{-h} \p{\frac{\pd \bar z'_1}{\pd \bar z_1}}^{-\bar h}e^{(h+\bar h)(\omega(z_1)-\omega(z'_1))}\times\dotsb\times \Psi$. The space of these transformations is three dimensional and the Lie algebra is spanned by
\al{
	L_3^{\{h_i,\bar h_i\}}&=\sum_i z_i\partial_i-\bar z_i\bar\partial_i+h_i-\bar h_i\\
	L_+^{\{h_i,\bar h_i\}}&=\sum_i z^2_i\partial_i +\bar\partial_i +\p{h_i-\bar h_i} z_i \\
	L_-^{\{h_i,\bar h_i\}}&=-\bar z^2\bar\partial -\partial+\p{h_i-\bar h_i}\bar z_i \ .
}
The above operators obey the usual angular momentum algebra, so it is sufficient to prove that any pair of these annihilate the correlation function. The operator $L_3^{\{h_i,\bar h_i\}}$ simply add the total power of all $z_i$'s and subtract the power of all $\bar z_i$'s, and subtract the total spin of all operators. So getting the $L_3^{\{h_i,\bar h_i\}}$ eigenvalue of  $\Psi(\{\mathbf x_i\})$ is a matter of power counting. We get that the requirement that $\Psi(\{\mathbf x_i\})$ is a zero eigenstate of $L_3^{\{h_i,\bar h_i\}}$ amounts to
\ml{
	\sum_{\alpha<\beta}\p{\mathbf r_\alpha^T \mathbf r_{\beta}-\mathbf{\bar r}^T_\alpha \mathbf{\bar r}_{\beta}}\\
	-\sum_\alpha\p{\sigma_\alpha-\bar\sigma_{\alpha}}+\sum_\alpha\p{h_\alpha-\bar h_\alpha}=0 \ .
}
Inserting the expressions \eqref{conformalweight} we get
\ml{
	\frac12\sum_{\alpha\beta}\p{\mathbf r_\alpha^T\mathbf r_{\beta}-\mathbf{\bar r}_\alpha^T \mathbf{\bar r}_{\beta}}-\frac{N_\phi}2\sum_\alpha\p{ \mathbf r_\alpha^T\mathbf c -\mathbf {\bar r}^T_\alpha \mathbf {\bar c}}\\
	-\sum_\alpha\p{ \mathbf r_\alpha^T \boldsymbol \Sigma -\mathbf {\bar r}_\alpha^T\boldsymbol {\bar \Sigma}}=0 \ .
}
This equation is implied by the neutrality condition, which is seen by subtracting \eqref{neutrality1} from \eqref{neutrality2}, multiplying with $\frac12r^T_\beta$, and summing over $\beta$.

It is now sufficient to prove that $L_-^{\{h_i,\bar h_i\}}$ annihilate the correlator. The commutator of $L_-^{\{h_i,\bar h_i\}}$ with the factors $z_i-z_j$ and $\pd_\alpha+\frac{C\bar z_\alpha}{1+z_\alpha\bar z_\alpha}$ is zero, and the commutator with $(1+z\bar z)^{-1}$ is $\bar z(1+z\bar z)^{-1}$. So we can commutate the $L_-^{\{h_i,\bar h_i\}}$ through the chiral part of the correlator, at the cost of changing the coefficient $\p{h_i-\bar h_i}$ in front of $\bar z$ in $L_-^{\{h_i,\bar h_i\}}$. Finding the coefficient is a matter of counting the factors $(1+z\bar z)^{-1}$. Using the neutrality condition again we see that $L_-^{\{h_i,\bar h_i\}}$ turn into $L_-^{\{0,\bar h_i\}}$ when commutating it through the chiral part of the correlator. So we are left with proving that $L_-^{\{0,\bar h_i\}}$ annihilate the anti-chiral part. Using the commutation relations we see that we instead can show that $L_3^{\{0,\bar h_i\}}$ and $L_+^{\{0,\bar h_i\}}$ annihilate the anti-chiral part. To prove that $L_3^{\{0,\bar h_i\}}$ does this, is, as before, just a matter of power counting, and $L_+^{\{0,\bar h_i\}}$ turns into $L_+^{\{0,0\}}$ when commutated through the anti-chiral part, analogous to what happened with $L_-$ when moving through the chiral part. So we have proved that the operators 	$L_3^{\{h_i,\bar h_i\}}$, $L_+^{\{h_i,\bar h_i\}}$ and $L_-^{\{h_i,\bar h_i\}}$ annihilate the correlator.

\subsection{Non-isometric conformal transformations}
In addition to the rotations there are three additional linear independent conformal transformations. With the regularization we have chosen the non-primary fields, \ie all which contain derivatives, will not be conformal invariant. The primary fields however will be invariant, under all analytic transformations, not just the Möbius transformations. To see this one just has to notice that the factors $(1+z\bar z)$ will transform such that they cancel $e^{2(\omega(z')-\omega(z))}$, and the rest of the correlation function is the same as in planar geometry. We also see that if the derivatives in the quasi-primary fields only would have acted on the zero modes of $L_3$ -- \ie if we in the regularization would have pointsplitted by a rotation around $\mathbf x=0$ -- then the same argument would have been true also for the quasi-primary fields, but then correlation function would not have been normalizable. We could have kept the conformal invariance, but then we would have needed to write the coherent state wave function not as a single correlator but as several correlators with different placements of the derivatives in such a way that the non-normalizable parts would cancel. 

\section{Convolution with the coherent state kernel}
\label{app:projection}
	To get the wave function in position basis $\Psi(\{\xi_i\})$ we must convolute with the coherent state kernel
	\al{
		\mathcal A\frac{(N_\phi+1)(1+\xi\bar z)^{N_\phi}}{ (1+z\bar{z})^{N_\phi/2}(1+\xi\bar{\xi})^{N_\phi/2}} \ .
	}
	We have
	\ml{
		(z\pd-(n+m-1+N_\phi/2))\times\dotsb\\
		\dotsb\times(z\pd-(n+N_\phi/2))\pd^n \braket{\xi|z}
		\propto \frac{\bar z^n}{(1+z\bar z)^m}\braket{\xi|z} \ ,
	}
	so the convolution can be done algebraically by replacing the factors not on LLL form, \ie, $z^a(1+z\bar z)^{-N_\phi/2}$, with the differential operators one gets when partially integrating the derivatives above, so they act on the coherent state wave function. 
	
	We also notice that if we had a total derivative on a purely chiral wave function, \ie, $\pd-\frac{N_\phi\bar z}{2(1+z\bar z)}$, then, when convoluting, the derivative would cancel against the factor $\frac{N_\phi\bar z}{2(1+z\bar z)}$ and the wave function would vanish. This is the reason for why the convention from Refs. \onlinecite{hermanns08}-\onlinecite{suorsa11b}, to only use overall derivatives, cannot be used on the sphere.
	
\bibliography{qHwavefunction}

\end{document}